\documentclass[12pt,preprint]{aastex}

\usepackage{emulateapj5}

\def\ph{\phantom{m}}

\shorttitle{THE MOMENT OF INERTIA OF J0737-3039A}
\shortauthors{MORRISON, BAUMGARTE, SHAPIRO \& PANDHARIPANDE}

\begin{document}

\title{The Moment of Inertia of the Binary Pulsar J0737-3039A:
Constraining the Nuclear Equation of State}
\author{I.~A.~Morrison\altaffilmark{1},
T.~W.~Baumgarte\altaffilmark{1,2,3},
S.~L.~Shapiro\altaffilmark{2,4} and
V.~R.~Pandharipande\altaffilmark{2}}

\altaffiltext{1}{Department of Physics and Astronomy, Bowdoin College,
Brunswick, ME 04011}

\altaffiltext{2}{Department of Physics, University of Illinois at
Urbana-Champaign, Urbana, IL 61801}

\altaffiltext{3}{Fellow of the J.~S.~Guggenheim Memorial Foundation}

\altaffiltext{4}{Department of Astronomy and NCSA, University of
Illinois at Urbana-Champaign, Urbana, IL 61801}

\begin{abstract}
We construct numerical models of the newly discovered binary pulsar
J0737-3039A, both with a fully relativistic, uniformly rotating,
equilibrium code that handles arbitrary spins and in the relativistic,
slow-rotation approximation.  We compare results for a representative
sample of viable nuclear equations of state (EOS) that span three,
qualitatively different, classes of models for the description of
nuclear matter. A future dynamical measurement of the neutron star's
moment of inertia from pulsar timing data will impose significant
constraints on the nuclear EOS.  Even a moderately accurate
measurement ($\lesssim 10 \%$) may be able to rule out some of these
competing classes.  Using the measured mass, spin and moment of
inertia to identify the optimal model computed from different EOSs,
one can determine the pulsar's radius.
\end{abstract}


\section{Introduction}
\label{sec:intro}

The recently discovered binary neutron star system J0737-3039 has many
very interesting and exciting properties
\citep[e.g.][]{betal03,letal04}.  Both stars in J0737-3039 have been
identified as pulsars, the binary is about ten times closer to the
Earth than the Hulse-Taylor binary B1913+16, and the binary separation
is significantly less than that of B1913+16 \citep{betal03}.  As a
consequence, relativistic effects affecting the orbital dynamics play
a much larger role in J0737-3039 than in all previously discovered
binary neutron stars.  One such effect that may become measurable soon
from pulsar timing data is spin-orbit coupling \citep{ds88,w95}.  In
particular, Lens-Thirring precession will yield the moment of inertia
of the more rapidly spinning pulsar J0737-3039A \citep{letal04,o04}.
Such a determination would constitute the first time that the moment
of inertia of a neutron star (or any other star) has been measured by
dynamical means. The recently launched Gravity Probe B experiment will
measure Lens-Thirring as well as geodetic precession of a gyroscope in
orbit about the Earth. Geodetic precession has been measured in the
binary pulsar system B1534+12 \citep{sta04}.

Since the star's gravitational mass and spin frequency are well
determined, a measurement of the moment of inertia could provide a
stringent constraint on the nuclear equation of state (EOS).  Using
the measured mass, spin and moment of inertia to select the optimal
candidate from a suite of theoretical models constructed from
competing EOSs, one can also determine the pulsar's radius.

In this paper we point out that even a measurement of the moment of
inertia of modest accuracy ($\lesssim 10\%$) may help differentiate
between competing approaches that are presently adopted for the
description of nuclear matter and the construction of nuclear EOSs
\citep[for a general review of modern EOSs see][]{lp00,lp01}. In
Section \ref{sec:eos} we briefly review three qualitatively distinct
classes of EOSs; in Section \ref{sec:moi} we adopt representative EOSs
in each of these classes to construct numerical models of J0737-3039A
and determine its moment of inertia; in Section \ref{sec:disc} we
briefly discuss our findings.


\section{Equations of State}
\label{sec:eos}

In this paper we consider six different nuclear EOSs.  They are chosen
to represent three different classes of EOSs, distinguished by the
formalism used in their construction (see discussion below).  All six
of these EOSs give maximum gravitational masses that exceed the mass
of the most massive neutron star \citep[PSR J1518+4904,
$M/M_{\odot}=1.56$;][]{tc99} residing in the six known binary radio
pulsar systems.  These systems provide neutron stars with the most
accurately determined masses.  Some pulsar masses determined with
greater uncertainty (e.g., PSR J0751+1807, believed to have a low-mass
white dwarf companion and an estimated mass $M/M_{\odot} \sim 2.2$;
see, e.g., Table 1 in Stairs 2004) are greater than the maximum
allowed masses for some of the chosen six EOSs (see Table 1 below).
If confirmed, these objects would rule out these particular EOSs, or
require adjustment of some of the parameters on which they are based.

The first class of EOSs that we consider are results of
nonrelativistic many-body calculations using realistic models of two-
and three-nucleon interactions.  The two-nuclon interactions are
obtained by fitting observed nucleon-nucleon scattering data, and the
three-nucleon interactions are based on the energies of light nuclei
and/or properties of symmeric nuclear matter (SNM) with comparable
number of neutrons and protons.  To represent these class I EOSs we
choose two examples. The first is APR (Akmal, Pandharipande, \&
Ravenhall 1998), which adopts a modern two-nuclon interaction (A18),
together with its relativistic boost correction and the UIX
three-nucleon potential. The second is FPS (Lorenz, Ravenhall, \&
Pethick 1993), which is based on an older two-nucleon interaction
(U14), and phenomenological density dependent terms to represent the
effects of three-nucleon forces.  The main difference between these
two EOSs is that FPS exactly reproduces the empirical properties of
SNM, but not light nuclei, while the A18+UIX forces used by APR
reproduce both the binding energies of light nuclei and SNM properties
fairly accurately.

An important property of the EOSs belonging to class I is that the
pressure of pure neutron matter (PNM) at desities in the
$10^{11}-10^{14}~\mbox{g~cm}^{-3}$ range is close to half of that of
noninteracting neutron gas (Carlson, Morales, Pandharipande, \&
Ravenhall 2003).  This follows from the known, large neutron-neutron
scattering length, $a \sim 18$ fm, which makes the dimensionless
parameter $ak_F$, where $k_F$ is the Fermi momentum of PNM, larger
than unity, and results in relatively smaller crusts, radii and
moments of inertia for neutron stars with $M < 1.6 M_{\odot}$.
 
The second, class II, of EOSs are based on relativistic mean-field
approximations.  A mean-field Lagrangian is chosen to reproduce the
empirical properties of SNM, and it is used to calculate the EOS of
neutron star matter, which is closer to PNM, using the mean-field
approximation.  SNM is charged, and in absence of Coulomb forces it is
believed to be bound with an energy of $\sim - 16$ MeV per nucleon,
and to have zero pressure at a density of $\sim 2.5 \times
10^{14}~\mbox{g~cm}^{-3}$.  At lower densities its pressure is
negative and it becomes unstable at densities less than $\sim 1.5
\times 10^{14}~\mbox{g~cm}^{-3}$.  On the other hand PNM has possitive
energy and pressure at all densities as far as we know.  Thus it is
most challenging to extrapolate the properties of SNM to those of PNM.
The relativistic mean field Lagrangian contains no information on
neutron-neutron scattering, such as the large value of the scattering
length $a$, and therefore the pressure of class II EOSs approaches
that of noninteracting neutron gas at desities less than
$10^{14}~\mbox{g~cm}^{-3}$.  They yield neutron stars with relatively
larger crusts, radii and moments of inertia.  We choose MS1 (Muller and
Serot 1996) and GM3 (Glendenning \& Moszkowski 1991) to represent
class II.

The class III of EOSs we consider assume that neutron stars are
entirely made up of strange quark matter.  This hypothetical form of
matter is electrically neutral and has zero pressure at densities of
order $10^{14}~\mbox{g~cm}^{-3}$.  There is no information or
observational evidence for this matter.  We consider two examples:
SQM1 and SQM3 (Prakash, Cooke, \& Lattimer 1995).

Tables of the comoving total mass density $\rho$, the rest mass
density $\rho_0$ and the pressure $P$ of the chosen EOSs were kindly
provided by M. Prakash and D. G. Ravenhall.  To improve the
convergence of our numerical codes we found it helpful to increase the
number of data points using third-order interpolation between the raw
data.


\section{The Moment of Inertia}
\label{sec:moi}

Adopting the EOSs discussed in Section \ref{sec:eos} we compute the
moment of inertia of the pulsar J0737-3039A by performing two independent
relativistic calculations. We assume that the pulsar is
rotating {\it uniformly}; while nascent neutron stars can form with
appreciable differential rotation, viscosity and magnetic fields will
combine to drive them to uniform rotation  on a timescale much
shorter than the age of the pulsar inferred from its  spindown, 210 Myr.
In addition, the precision of the underlying
pulsar clock mechanism strongly argues in favor of solid-body rotation.

In the first calculation, we adopt the numerical code of Cook, Shapiro
\& Teukolsky 1994, hereafter CST, to construct stationary models of
uniformly rotating neutron stars. This code treats arbitrary spin
rates without approximation. We use it to compute models that have the
gravitational mass $M = 1.337 M_{\odot}$ and the angular velocity
$\Omega = 276.8 \mbox{s}^{-1}$ (corresponding to a spin period $P =
27.7$ ms) of J0737-3039A.  We refer to CST for details of the code,
including how the code utilizes tabulated EOSs.  The code computes the
angular momentum $J$
\begin{equation} \label{eq:J_SS}
J = \int T^\mu_{\ph \nu} \xi_{(\phi)}^{\nu} d^3 \Sigma_\mu
\end{equation}
where $T^{\mu\nu}$ is the stress-energy tensor for a perfect fluid,
\begin{equation} \label{eq:T}
T^{\mu\nu} = (\rho + P)u^\mu u^\nu + P g^{\mu\nu},
\end{equation}
and $\xi_{(\phi)}^{\nu}$ is the Killing vector in the azimuthal direction
reflecting axisymmetry.  In (\ref{eq:T})
$u^{\mu}$ is the fluid four-velocity, which for uniformly rotating,
axisymmetric stars takes the form
$
u^{\mu} = u^t (1,0,0,\Omega)
$
where $\Omega = u^{\phi}/u^t$ is the stellar angular velocity.
The definition (\ref{eq:J_SS}) then reduces to
\begin{equation}
J = \int (\rho + p) u^t (g_{\phi \phi}u^\phi + g_{\phi t} u^t)
\sqrt{-g} d^3 x \label{eq:J2}
\end{equation}
where $g \equiv \rm{det}(g_{\alpha\beta})$.  From the angular momentum
$J$ the moment of inertia is determined as $I \equiv J/\Omega$.

A simple estimate shows that the stellar frequency $\Omega = 276.8$ Hz
is much less than the Kepler frequency at the stellar equator,
$\Omega_{\rm max} \approx (G M/R^3)^{1/2}$.  For all the EOSs we
consider here $\Omega/\Omega_{\rm max} < 0.035$.  This implies that
deviations from sphericity are very small, so that the moment of
inertia can also be approximated from spherical stellar models.  Our
second calculation therefore adopts this slow-rotation approximation
\citep{h67,cm74,gw92,rp94} which we briefly review as follows.

\begin{table*}
\caption{Numerical results for J0737-3039A ($M = 1.337 M_{\odot}$ and
$\Omega = 276.8 \mbox{s}^{-1}$) in the slow-rotation approximation}
\centerline{
\begin{tabular}{ccccccccccc}
\hline \hline
Class & EOS &
$M_{\rm max}/M_{\odot}$~\tablenotemark{a} &
$J$~\tablenotemark{b} &
$I$~\tablenotemark{c} &
$I/MR^2 $ &
$\rho_c$~\tablenotemark{d} &
$M_0/M_{\odot}$~\tablenotemark{e} &
$M^c_0/M_{\odot}$~\tablenotemark{f} &
$R$~\tablenotemark{g} &
$GM/c^2 R$ \\
\hline
I & APR~\tablenotemark{h} & 2.20 & 3.42 & 1.24 & 0.348 & 9.61 & 1.47 & 1.37  & 11.56 & 0.1707 \\
  & FPS~\tablenotemark{i} & 1.80 & 3.14 & 1.14 & 0.358 & 12.1 & 1.48 & 1.37  & 10.92 & 0.1809  \\
\hline
II & MS1~\tablenotemark{j} & 1.82 & 4.61 & 1.66 & 0.326 & 6.81 & 1.45 & 1.34 & 13.87 & 0.1424  \\
   & GM3~\tablenotemark{k} & 1.56 & 4.14 & 1.49 & 0.325 & 9.10 & 1.46 & 1.35 & 13.12 & 0.1502  \\
\hline
III & SQM1~\tablenotemark{l} & 1.56 & 2.72 & 0.982& 0.474 & 15.1 & 1.55 & 1.44 & 8.830 & 0.2236 \\
    & SQM3~\tablenotemark{l} & 1.99 & 3.91 & 1.41 & 0.459 & 7.00 & 1.50 & 1.40 & 10.76 & 0.1836 \\
\hline
\end{tabular}}
\tablenotetext{a}{~maximum allowed gravitational mass for non-rotating star}
\tablenotetext{b}{~total angular momentum $(10^{47}\mbox{g cm}^2\mbox{s}^{-1})$, eqn. (\ref{eq:J2})}
\tablenotetext{c}{~moment of inertia $(10^{45}\mbox{g cm}^2)$, eqn. (\ref{eq:I})}
\tablenotetext{d}{~central mass-energy density $(10^{14}\mbox{g cm}^{-3})$}
\tablenotetext{e}{~rest mass}
\tablenotetext{f}{~rest mass of companion, J0737-3039B
($M = 1.25 M_{\odot}$, $\Omega = 2.27 \mbox{s}^{-1}$)}
\tablenotetext{g}{~circumferential radius $(\mbox{km})$}
\tablenotetext{h}{~A18 +$\delta v$ + UIX$^{*}$; \cite{apr98}} 
\tablenotetext{i}{~UV14 $+$ TNI; \cite{lpr93}}                
\tablenotetext{j}{~Field Theoretical; \cite{gm91}}            
\tablenotetext{k}{~Field Theoretical; \cite{ms96}}            
\tablenotetext{l}{~Quark Matter; \cite{pcl95}}                
\label{Table1}
\end{table*}

In the slow-rotation limit in spherical polar coordinates
the metric can be written as ($G=c=1$)
\begin{eqnarray} \label{eq:metric}
ds^2 & = & - e^{2\Phi} dt^2 + \left( 1 - \frac{2m}{r}
\right)^{-1} dr^2 \nonumber \\
& & - 2\omega r^2 \sin^2 \theta dt
d\phi + r^2 (d \theta^2 + \sin^2 \theta d \phi^2)
\end{eqnarray}
where $m(r)$ is the total mass within radius $r$
\begin{equation} \label{eq:mass}
\frac{dm}{dr} = 4 \pi r^2 \rho
\end{equation}
and where $\omega(r) \equiv \left. d\phi/dt \right|_{\rm ZAMO}$ is the
Lense-Thirring angular velocity of a zero-angular momentum observer
(ZAMO) with $p_{\phi} = 0$.  Up to first order in $\omega$ all metric
potentials are spherically symmetric and only depend on the areal
radius $r$.  In the stellar interior, Einstein's equations reduce to
\begin{equation} \label{eq:dphi}
\frac{d\Phi}{dr} = \frac{m + (4\pi r^3 p)}{1 - 2m/r} \;\; (r < R),
\end{equation}
and
\begin{equation} \label{eq:domg}
\frac{1}{r^3} \frac{d}{dr} \left( r^4 j \frac{d\bar{\omega}}{dr} \right)
+ 4 \frac{dj}{dr} \bar{\omega} = 0 \;\; (r < R),
\end{equation}
where following convention we define $\bar{\omega} \equiv \Omega -
\omega$ and
\begin{equation} \label{eq:defj}
j \equiv \left( 1 - \frac{2m}{r} \right)^{1/2}\; e^{-\Phi}.
\end{equation}
The stellar structure is determined by the Tolman-Oppenheimer-Volkoff
eq.~\citep{ab76}
\footnote{Note that in eq.~(3.11) of \citeauthor{ab76}
this expression has an incorrect exponent}
\begin{equation} \label{eq:tov}
\frac{dP}{dr} = - (\rho + P) \frac{m + 4 \pi r^3 P}{1 - 2m/r}.
\end{equation}

Exterior to the stellar surface at $r=R$, defined as the radius $R$ at
which the pressure first vanishes $P(R) = 0$, the metric reduces to
\begin{eqnarray} \label{eq:phiExt}
e^{2 \Phi} = \left(1 - \frac{2M}{r}\right)\;\;\; (r > R),
\end{eqnarray}
and
\begin{equation} \label{eq:omgExt}
\omega = \frac{2J}{r^3}
\;\;\;\;\;\;\;\; (r > R)
\end{equation}
where $M = m(R)$ is the total gravitational mass.

On the surface of the star the interior and exterior solutions
are matched by satisfying the boundary conditions
\begin{equation} \label{eq:bc-omg}
\bar{\omega}(R) = \Omega - \frac{R}{3} \frac{d\bar{\omega}}{dr},
\end{equation}
and
\begin{equation} \label{eq:bc-phi}
\Phi(R) = \frac{1}{2} \ln(1 - 2M/R).
\end{equation}
In practice this is accomplished by choosing arbitrary initial values
$\Phi_c$ and $\omega_c$ at $r=0$ for the integration of
(\ref{eq:dphi}) and (\ref{eq:domg}) to the surface $r=R$. The boundary
conditions are then met by multiplying the interior solutions by
appropriate constants.

The moment of inertia $I= J/\Omega$ can then be computed from
eq.~(\ref{eq:J2}).  Using $\Omega = u^{\phi}/u^t$ and keeping only
terms up to first order in $\Omega$ and $\omega$ we find
\begin{equation} \label{eq:I}
I  \approx \frac{8 \pi}{3} \int^R_0 \frac{(\rho + p) e^{-\Phi}}
{(1 - 2m/r)^{1/2}} \frac{\bar{\omega}}{\Omega} r^4 dr.
\end{equation}
This slow-rotation approximation for the moment of inertia neglects
deviations from sphericity and is independent of the angular velocity
$\Omega$.  

In this framework, the moment of inertia can be computed by solving
ordinary differential equations, namely the Tolman-Oppenheimer-Volkoff
eqs.~(\ref{eq:mass}) and (\ref{eq:tov}) together with
eqs.~(\ref{eq:dphi}) and (\ref{eq:domg}), followed by a numerical
quadrature, (\ref{eq:I}).  This integration can be carried out with
almost arbitrary accuracy, so that that the error is dominated by the
slow-rotation approximation.  From $(\Omega/\Omega_{\rm max})^2 \sim
10^{-3}$ we expect this error to be in the order of about 0.1 \%
(which we have verified by comparison with our calculations using the
CST code).  For most EOSs our datasets are fairly sparse, which leads
to an interpolation error that is closer to about 1 \% for some EOSs,
far larger than the intrinsic error arising from the slow-rotation
approximation.  We summarize our numerical results in Table
\ref{Table1}, constructed using this approximation.  All results
obtained with the CST code were within 1 \% of the results in Table
\ref{Table1}, the difference dominated by interpolation error within
the EOS datasets, which is carried out differently in the two
calculations. (We found closer agreement for those EOSs for which more
datapoints are available).  We also point out that the CST code
calculations showed that the oblateness of the stars is very small for
all EOSs $R_e/R_p - 1 \lesssim 10^{-3}$, where $R_e$ and $R_p$ are the
equatorial and polar radius.


\section{Discussion}
\label{sec:disc}

It is evident from Table \ref{Table1} that the moment of inertia $I$
for a neutron star like J0737-3039A depends quite strongly on the EOS.
Hence a measurement of $I$ could impose new constraints on the EOS,
refining some of the parameters that enter the models of nuclear
matter, or even ruling out whole classes of models that currently
exist.

Within each Class I and II, the predictions for the neutron star
radius and moment of inertia are fairly similar, but between these two
classes, the values are quite different. Class I predicts a radius
$\lesssim 12$ km and a moment of inertia $\lesssim 1.2\times 10^{45}
\mbox{g cm}^2$, while Group II predicts a radius of about $\gtrsim 13$
km and a moment of inertia $\gtrsim 1.5 \times 10^{45}\mbox{g cm}^2$.
This implies that an observational measurement of $I$ even with a
modest accuracy (say 10 \% or so) may distinguish these different
groups, establishing whether ``realistic'' many-body nuclear models
based on nuclear matter and experimental data (Class I) or mean-field
models (Class II) provide better descriptions of neutron star matter.

There is a much larger spread in our results for Class III,
which seem to depend quite strongly on 
various free parameters in the EOS.  These parameters affect the
(surface) density at which the pressure vanishes, and hence the radius
and moment of inertia of the stellar models.  A measurement of $I$ may
serve to constrain these parameters.

As an aside we point out that the pulsar's companion, J0737-3039B,
which is also a pulsar, has a surprisingly small gravitational mass of
only $M = 1.25M_{\odot}$.  The corresponding baryonic rest mass
depends on the EOS and ranges from 1.34 $M_{\odot}$ to 1.37
$M_{\odot}$ for Classes I and II \citep[see Table \ref{Table1}; also
Fig.~3 in][]{mbs04}.  Most supernova simulations lead to remnants that
are more massive than this \citep[e.g.][]{whw02}, especially if
general relativistic effects are included \citep[e.g.][]{f99}.  It is
likely that J0737-3039B originated from a close helium star-neutron
star binary \citep{dh04,wk04}, which will clearly affect the dynamics
of the supernova explosion and hence the remnant mass.  It is
therefore too early to draw conclusions, but any future simulations of
supernovae in close binaries ought to be able to produce such small
mass remnants.  That outcome may be possible for some EOSs but not for
others, which would put a further constraint on the nuclear EOS.

\acknowledgments

It is a pleasure to thank A.~Burrows, C.~Fryer, H.-Th.~Janka, and
F.~Rasio for useful conversations, and M.~Prakash and D.~Ravenhall for
providing data for some of the EOSs.  This paper was supported in part
by NSF Grant PHY01-39907 at Bowdoin College and NSF Grants
PHY02-05155, PHY03-45151, and PHY03-55014 and NASA Grant NNG04GK54G at
the University of Illinois at Urbana-Champaign.


\end{document}